\documentclass[]{jfm}

\usepackage[dvipsnames]{xcolor}
\usepackage{graphicx}
\usepackage{stackengine}
\usepackage{newtxtext} 
\usepackage{newtxmath}
\usepackage{natbib}
\usepackage{soul}
\usepackage[hoptionsi]{overpic}
\usepackage{subcaption}
\usepackage{comment}
\usepackage{hyperref}
\hypersetup{
    colorlinks = true,
    urlcolor   = blue,
    citecolor  = blue,
    linkcolor=blue,
}

\newcommand{\RomanNumeralCaps}[1]
\linenumbers

\newcommand{\black}{\color{black}} 
\newcommand{\red}{\color{black}}

\title{Modelling turbulence in axisymmetric wakes: an application to wind turbine wakes}

\author{Majid Bastankhah\aff{1}
  \corresp{\email{majid.bastankhah@durham.ac.uk}},
   Jenna K. Zunder \aff{1},
   Peter E. Hydon \aff{2},
 Charles Deebank\aff{3}
 \and Marco Placidi\aff{3}}

\affiliation{\aff{1}Department of Engineering, Durham University, Durham DH1 3LE, UK
\aff{2}School of Mathematics, Statistics and Actuarial Science, University of Kent, Canterbury CT2 7NF, UK
\aff{3} EnFlo Research Centre, University of Surrey, Guildford, GU2 7XH, UK}

\begin{document}
\maketitle

\begin{abstract}
A novel fast-running model is developed to predict the three-dimensional (3D) distribution of turbulent kinetic energy (TKE) in axisymmetric wake flows. This is achieved by mathematically solving the partial differential equation of the TKE transport using the Green's function method. The developed solution reduces to a double integral that can be computed numerically for a wake prescribed by any arbitrary velocity profile. It is shown that the solution can be further simplified to a single integral for wakes with Gaussian-like velocity-deficit profiles. \black Wind tunnel experiments were performed to compare model results against detailed 3D laser Doppler anemometry data measured within the wake flow of \black a porous disk subject to a uniform freestream flow. \black Furthermore, the new model is \black used to estimate the TKE distribution at the hub-height level of the rotating non-axisymmetric wake of a model wind turbine immersed in a rough-wall boundary layer. \black Our results show the \red important \black impact of \red operating conditions \black on TKE generation in wake flows, an effect not fully captured by existing empirical models. 
 \black The wind-tunnel data also provide insights into the evolution of important turbulent flow quantities such as turbulent viscosity, mixing length, and the TKE dissipation rate in wake flows. \black Both mixing length and turbulent viscosity are found to increase with the streamwise distance. The turbulent viscosity however reaches a plateau in the far-wake region. Consistent with the non-equilibrium theory, it is also observed that the normalised energy dissipation rate is not constant, and it increases with the streamwise distance.\black
\end{abstract}


\section{Introduction}
\label{sec:intro}

The importance of axisymmetric wake flows lies in their pivotal role in optimising the efficiency and environmental impact of various engineering systems, ranging from wind energy to aerospace design and pollution control. The study of turbulence in axisymmetric wake flows at high Reynolds numbers is a fundamental fluid-mechanics problem treated in many turbulence textbooks \citep[e.g.][]{pope2000}. It is also practically significant because it helps researchers predict and mitigate issues related to drag, flow instability, heat exchange, and energy efficiency. For instance, turbulence in wind turbine wakes serves as a double-edged sword. It aids the wake recovery, thus providing more energy for downwind turbines in a wind farm, while simultaneously amplifying the fatigue loads experienced by those very turbines \citep{stevens2017review}.

The need for fast-running models to predict wake flows such as those of wind turbines is essential, \black as high- fidelity computational fluid dynamics (CFD) modeling  or experiments are still too expensive/time-consuming and are, therefore, impractical for the optimisation and real-time control of these flows \citep{meneveau2019}. Typically, CFD modelling entails choosing a computational discretisation technique, creating often complex numerical grids, selecting a suitable turbulence model, and conducting post-processing on the data \citep{meneveau2019}. This demands considerable manpower and expertise in fluid mechanics (and turbulence modelling), which typically is not always readily available in industries such as wind energy due to its multidisciplinary nature and diverse workforce. Hence, engineering wake models are often preferred over high-fidelity CFD models in industrial applications. \black While a substantial amount of research has been dedicated to developing fast-running models to predict mean velocity distribution in wind turbine wakes \citep[\textcolor{black}{see}][and references therein]{porte2020}, estimation of turbulence in these flows predominantly relies on purely empirical methods \citep[e.g.][]{crespo1996,ishihara2018new} owing to the intricate nature of turbulence modelling. \black Existing \black analytical \black engineering wind-farm models \citep[e.g.][]{Niayifar2016,bastankhah2021cumulative_wake,zong2020momentum,lanzilao2022new} typically determine the wake recovery rate based on the incoming turbulence level for each turbine. The TKE generated by upwind turbines, predicted based on empirical models such as \cite{crespo1996}, then serves as initial conditions for predicting the wake recovery rate of downwind turbines. This emphasises the importance of advancing physics-based methods to predict the TKE distribution. \black To the best of our knowledge, \black this work is the first attempt to develop a physics-based engineering TKE model for wake flows\black. The model derivation is elaborated in \S\,\ref{sec:model}, the experimental setup \black for two cases \black used to validate model predictions is described in \S\,\ref{sec:exp}. Results are then discussed in \S\,\ref{sec:Res}, before a summary is provided in \S\,\ref{sec:summary}.\vspace{- 5 mm}
\section{Mathematical model development}\label{sec:model}
The axisymmetric wake is described using a cylindrical coordinate system with $(x,r,\theta)$, where $x$ is the streamwise distance from the object causing the turbulent wake, $r$ is the radial distance from the centre of the wake, and $\theta$ is the azimuthal angle. 
Mean and fluctuating velocity components in the $(x,r,\theta)$ coordinate system are indicated by $(U,V,W)$ and $(u,v,w)$, respectively. 
Due to the assumption of axisymmetry, $\partial/\partial \theta=0$. 
 In the following, $<>$ denotes time averaging. The TKE denoted by $k$ is defined by $k=0.5\langle q^2\rangle$ where $q^2=u^2+v^2+w^2$. 
The steady-state TKE transport equation, neglecting \black pressure-velocity covariance \black and viscous diffusion but including swirl, reads as \citep{shiri2010turbulence}
 \begin{multline}\label{eq:TKE-transport1}
\underbrace{U\frac{\partial k}{\partial x}+\left[V\frac{\partial k}{\partial r}\right]}_{\text{advection}}+\underbrace{\left[\frac{\partial\langle uq\rangle}{2\partial x}\right]+\frac{1}{2r}\frac{\partial \left(r\langle vq \rangle\right)}{\partial r}}_{\text{diffusion}}+\underbrace{\varepsilon\vphantom{\left(\frac{2\pi}{365}\right)}}_{\text{dissipation}}=
    \underbrace{\left[\langle vw\rangle\frac{W}{r}\right]-\left[\langle w^2\rangle \frac{V}{r}\right]}_{\text{production}}\\\underbrace{-\left[\langle v^2\rangle \frac{\partial V}{\partial r}\right]-\left[\langle u^2\rangle \frac{\partial U}{\partial x}\right]-\left[\langle uv\rangle \frac{\partial V}{\partial x}\right]-\left[\langle uw\rangle \frac{\partial W}{\partial x}\right]-\left[\langle vw\rangle \frac{\partial W}{\partial r}\right] - \langle  uv\rangle \frac{\partial U}{\partial r}}_{\text{production}}.
\end{multline}
The following simplifications and assumptions are made to be able to mathematically solve \eqref{eq:TKE-transport1}. First, we model the diffusion (i.e. transport) terms based on the gradient-diffusion hypothesis \black given by \black \citep{pope2000}
\begin{equation}\label{eq:gradient-diffusion}
    \frac{\partial \langle uq\rangle}{2\partial x}+\frac{\partial (r\langle vq\rangle)}{2r\partial r}=-\frac{\partial}{\partial x}\left(\frac{\nu_T}{\sigma_k}\frac{\partial k}{\partial x}\right)-\frac{1}{r}\frac{\partial}{\partial r}\left( \frac{\nu_T}{\sigma_k}r\frac{\partial k}{\partial r} \right),
\end{equation}
  \black where $\nu_T$ is the turbulent viscosity and \color{black} $\sigma_k$ is the turbulent Prandtl number. The value of $\sigma_k$ generally depends on atmospheric stability \citep[see][among others]{li2019turbulent, basu2021turbulent}, with experimental results of $\sigma_k \in [0.7,0.92]$ for a neutral atmospheric boundary layer (ABL) flow \citep{businger1971flux, kays1994turbulent}. This work assumes $\sigma_k=1$ based on the Reynolds analogy, commonly used for turbulent flows \citep{pope2000}. \color{black} The validity of the gradient-diffusion hypothesis is \black further \black examined in \S\,\ref{sec:Res}. The terms in \eqref{eq:TKE-transport1} shown in square brackets are \black normally small compared to other terms for axisymmetric wake flows and can be thus neglected \citep{uberoi1970turbulent}. Neglecting these terms will be also supported by \black the budget analysis performed in \S\,\ref{sec:Res}\black. 
  
  Prior studies \citep[e.g.][]{wygnanski1970two, hussein1994} showed that $\nu_T$ is fairly uniform at \black the centre of axisymmetric wakes, but it decays at wake edges. While cross-stream variations of turbulent viscosity can be determined based on velocity profiles \citep{basset2022entrainment}, \black for simplicity, \black the common assumption of $\nu_T\approx\nu_T(x)$ is used herein \citep{pope2000}. Moreover, the dominant advection term (i.e. $U\partial k/\partial x$) is linearised by replacing $U$ with $U_0$, where the subscript $_0$ denotes the inflow. This approximation improves with distance from the origin, as the velocity deficit $\Delta U$ decreases. The TKE dissipation rate \black is written as \black
  \begin{equation}
      \varepsilon=C_\varepsilon k^{3/2}/l_m,
  \end{equation}
 where $l_m=l_m(x)$ is the mixing length. The normalised energy dissipation rate, $C_\varepsilon$, is traditionally assumed to be constant for high Reynolds number flows, but there has been a great deal of evidence in recent years showing that it may not be constant \citep[see the review of][and references therein]{vassilicos2015review}. 
 Therefore, we assume $C_\varepsilon=C_\varepsilon(x)$. Moreover, \black the turbulent viscosity is commonly modelled by \black 
  \begin{equation}
      \nu_T=ck^{1/2}l_m,
  \end{equation}
  where $c$ is a constant \citep{pope2000}. \black Hence, the TKE dissipation rate $\varepsilon$ can be expressed as \black
  \begin{equation}
      \varepsilon=\frac{\nu_T k}{\Psi(x)}, \qquad \textrm{where} \quad  \Psi(x)=\frac{cl_m^2(x)}{C_\varepsilon(x)}.
  \end{equation}
 Finally, the Boussinesq hypothesis is used to model the Reynolds shear stress, \black where \black
\begin{equation}
     \langle uv\rangle = -\nu_T \bigg( \frac{\partial U}{\partial r}+\frac{\partial V}{\partial x}\bigg).
\end{equation}
Note that $\partial V/\partial x\!<\!<\!\partial U/\partial r$, especially in the far-wake region, and thus it can be neglected. Given the hypotheses discussed above, \eqref{eq:TKE-transport1} can be reduced to
\begin{equation}\label{eq:TKE-transport2}
	\frac{U_0}{\nu_t(x)}\frac{\partial k_w(x,r)}{\partial x}-\frac{1}{r}\frac{\partial}{\partial r}\left( r\frac{\partial k_w(x,r)}{\partial r} \right) + \frac{1}{\Psi(x)}k_w(x,r) \approx \bigg(\frac{\partial U(x,r)}{\partial r}\bigg)^2.
\end{equation}
Note that in \eqref{eq:TKE-transport2}, the total TKE, $k$, is substituted with the wake-generated TKE, $k_w$, where $k_w=k-k_0$. 
This is only possible assuming that the spatial variations of $U_0$ and $k_0$ are negligible compared to variations caused by the wake. \black Moreover, the TKE dissipation rate in the background flow is assumed to be considerably smaller than the one in the wake\black. We therefore neglect the terms including $k_0$ on the left-hand side of \eqref{eq:TKE-transport2}, and $\partial U/ \partial r$ is only due to the wake-generated shear \black (i.e. $\partial U/\partial r=\partial U_w/\partial r$)\black. \black Note that the velocity profile $U(x,r)$ is assumed to be any smooth function with an arbitrary shape. \black   

The solution of the above equation is sought for the domain of $x\geq x_0$ and $r\geq 0$, where $x_0$ is the virtual origin. The initial and boundary conditions are defined as $k_w(x_0,r)=0$, $\partial k_w(x,0)/\partial r=0$, and $k_w(x,\infty)\rightarrow  0$ as $r\rightarrow \infty$, respectively. Due to the shear in the wake flow, the turbulence level normally starts increasing right from the origin of the wake, so we assume that $x_0=0$ is a good approximation. \black However, $x_0$ has been shown to be both positive and negative \citep{neunaber2022wind}, and, if included as a variable parameter, may improve wake model predictions \citep{neunaber2024leading}. Thus, $x_0$ \black has an arbitrary value in our model derivation for more flexibility. The solution involves two positive, monotonic functions of $x$ and a dummy variable $X$ (such that $x_0\leq X\leq x$), namely
\begin{equation}
	\phi(X,x)=\frac{1}{U_0}\int_{\xi=X}^x\nu_t(\xi) \,\mathrm{d}\xi,\qquad \psi(X,x)=\frac{1}{U_0}\int_{\xi=X}^x\frac{\nu_t(\xi)}{\Psi(\xi)} \,\mathrm{d}\xi,
\end{equation} 
where $\xi$ is a dummy variable. The exact solution of \eqref{eq:TKE-transport2}, achieved using the Green's function method is
\begin{equation}\label{eq:modsol}
	k_w(x,r)=\int\limits_{X=x_0}^x\int\limits_{\rho=0}^\infty\frac{\nu_t(X)}{2U_0\phi(X,x)}\,\exp\left\{-\,\frac{r^2+\rho^2}{4\phi(X,x)}\,-\psi(X,x)\right\}I_0\left(\frac{r\rho}{2\phi(X,x)}\right)\left(\frac{\partial U(X,\rho)}{\partial \rho}\right)^2\,\rho\,\mathrm{d}\rho\,\mathrm{d}X,
\end{equation}
\black
where $I_0$ is the modified Bessel function of the first kind\black, and $\rho$ is a dummy variable. See appendix \ref{appA} for more information on the derivation of \eqref{eq:modsol}. To compute the integrand in \eqref{eq:modsol}, the wake velocity profile $U(x,r)$ needs to be known. \black This model can be used with any \black axisymmetric \black velocity profile, however since Gaussian-type models are often used to represent the mean flow properties in wakes, these are presented here. A wake with a Gaussian velocity profile is given by \black \citep{tennekes1972first,Vermeulen1980,bastankhah2014new}
    \begin{equation}\label{gauss}
        U(x,r)=U_0\left[1-C(x)\textrm{exp}\left(-\frac{r^2}{2\sigma(x)^2}\right)\right],
    \end{equation}
where $\sigma(x)$ is the characteristic wake width at $x$, and $C(x)$ is the maximum normalised velocity deficit at each $x$. \black A double-Gaussian velocity profile is given by \citep{schreiber2020brief}
\begin{equation}\label{doublegauss}
    U(x,r)=U_0 \bigg[ 1- \frac{1}{2}C(x)\bigg(\exp\bigg(-\frac{(r-r_0)^2}{2 \sigma(x)^2}\bigg)+\exp\bigg(\frac{(r-r_0)^2}{2 \sigma(x)^2}\bigg) \bigg) \bigg],
\end{equation}
where $r_0$ is the radial position of the Gaussian extrema. For a wake with a single Gaussian velocity deficit profile as in \eqref{gauss}, \black one can simplify \eqref{eq:modsol} to
\begin{multline}\label{modsolgauss}
	k_w(x,r)=\int_{X=x_0}^x\frac{U_0\nu_t(X)C^2(x)}{(\sigma^2(X)+4\phi(X,x))^3}\,\left\{\sigma^2(X)r^2+4\phi(X,x)\left(\sigma^2(X)+4\phi(X,x)\right)\right\}\\
	\times\exp\left\{-\,\frac{r^2}{\sigma^2(X)+4\phi(X,x)}\,-\psi(X,x)\right\}\,\mathrm{d}X.
\end{multline}
For other wake flow profiles such as double-Gaussian, the double integral in \eqref{eq:modsol} cannot be reduced to a single integral, so both integrals need to be computed numerically. \black The reader is referred to Appendix \ref{appB} for a discussion on numerical integration of \eqref{eq:modsol} for an arbitrary velocity profile.\black

\begin{figure}
\begin{minipage}{0.75\textwidth}
\begin{subfigure}{\linewidth}
\topinset{(a)}{\includegraphics[width=1\linewidth]{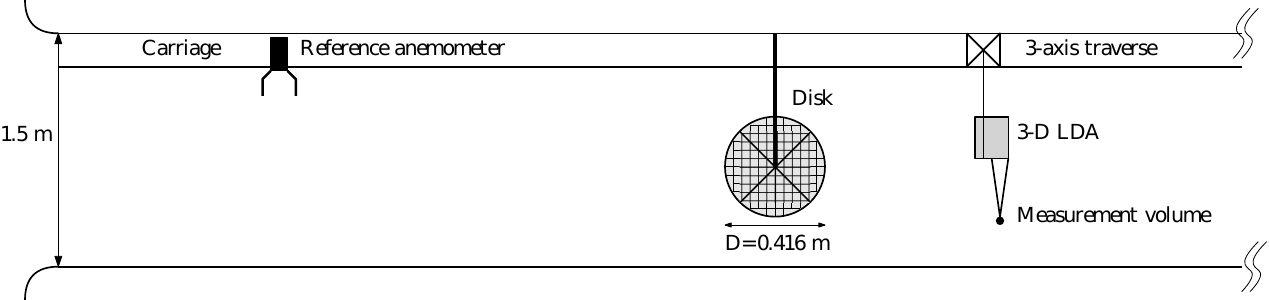}}{-0.12in}{-2.1in}
\end{subfigure}
\vspace{0.05\textwidth}
\begin{subfigure}{\linewidth}
\topinset{(b)}{\includegraphics[width=1\linewidth]{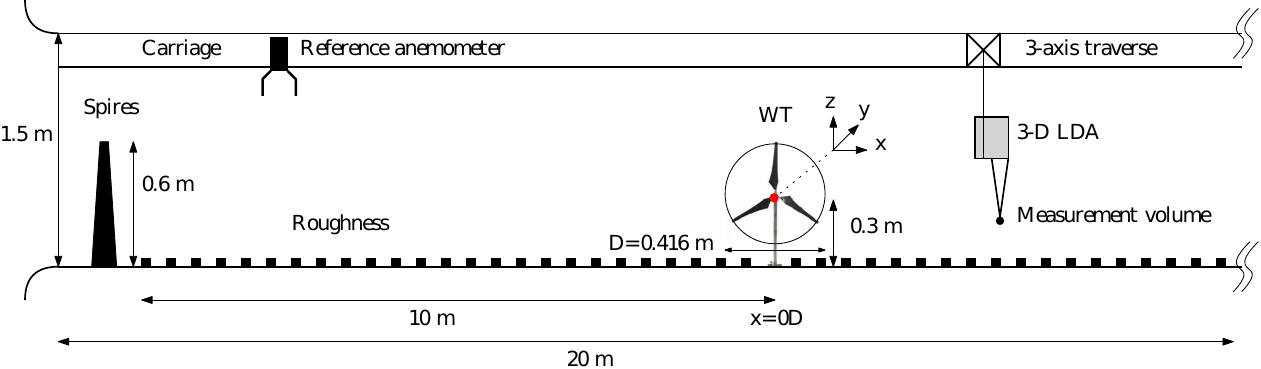}}{-0.12in}{-2.1in}
\end{subfigure}
\end{minipage}
\hspace{0.05\textwidth}
\begin{subfigure}{0.15\textwidth}
\topinset{(c)}{\includegraphics[width=1\linewidth]{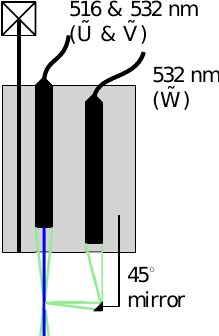}}{-0.12in}{-0.6in}
\end{subfigure}
\caption{\textcolor{black}{Sketch, and overall dimensions, of the wind tunnel setup: (a) disk in the freestream (FS) case, (b) wind turbine in the boundary-layer (BL) case, and (c) close-up of the 3D LDA setup  (not to scale). $\tilde{U}, \tilde{V}, \tilde{W}$ indicate the mean velocities in the Cartesian coordinate system.}}
\label{fig:WT}
\end{figure}

\section{Experimental setup}\label{sec:exp}
Experiments were conducted in the EnFlo wind tunnel at the University of Surrey with a test section of \textcolor{black}{20m $\times$ 3.5m $\times$ 1.5m (length $\times$ width $\times$ depth)}. 
\black Tests were run at the freestream speed 1.5 ms$^{-1}$, as measured by an ultrasonic anemometer mounted at the tunnel inlet \textcolor{black}{(see figure \ref{fig:WT})}. Two different canonical flows are considered in this work: (i) a porous disk in a uniform shear-free flow (i.e., a freestream (FS) case), and (ii) a turbine model in \black shear flow \black over a rough wall (i.e., a boundary-layer (BL) case). \textcolor{black}{These are depicted in figures \ref{fig:WT}(a) and (b), respectively.} For the free-stream case, a porous disk (of diameter $D=416$ mm) was manufactured out of a metallic grid arranged in an orthogonal coordinate system, with wires of diameter 1.2 mm and a square grid $3.8\times3.8$ mm$^2$ in size. The disk was mounted in the centre of the tunnel (both vertically and laterally) 
via a roof-mounted sting; \red its estimated thrust coefficient, $C_T$, is 0.65. \black
The mean incoming flow 
for the free-stream case was verified to be uniform within $0.39\%$ and characterised by a turbulence intensity of $TI=1/U_0\sqrt{1/3(\langle u^2_0\rangle+\langle v^2_0\rangle+\langle w^2_0\rangle})=1.98\%$.

For the boundary-layer case, a typical \black offshore boundary layer was achieved by a combination of 13 truncated triangular spires located at the tunnel inlet with roughness elements arranged in a staggered layout covering the entire tunnel floor. \textcolor{black}{The spires (flat plates) have the following dimensions: 60 mm (base), 4 mm (apex),
are 600 mm tall and spaced 266 mm in the tunnel's span. The roughness elements were arranged in a staggered layout covering the tunnel floor.} 
The values of $(u_*/U_{0})^2$ and roughness length $z_0$ are 2.2$\times 10^{-3}$ and 0.18 mm, respectively, where $u_*$ is the friction velocity. The boundary layer setup is depicted in figure \ref{fig:WT}(b), with further information provided in \cite{Placidi:2023b}. \textcolor{black}{The flow mean uniformity across the spanwise direction of the boundary layer at hub height $z_h$ is within $1.16\%$}. 
The wind turbine used in this \textcolor{black}{case} is a 1:300 scaled rotating model of a 5 MW offshore wind turbine with a rotor diameter $D$ of $416$ mm (matching the porous disk case) and a hub height $z_h$ of 300 mm. The blades are tapered and twisted flat plates to account for the much lower operating laboratory Reynolds number compared to the full-scale counterpart. See \cite{Hancock:2014b} and \cite{Placidi:2023b} for more information on turbine design and characteristics. The model tip-speed ratio $\lambda$ was kept at $6\pm1.5\%$, which resulted in an estimated $C_T$ of $0.48$, \textcolor{black}{originally based on wake measurements in uniform flow \citep{Hancock:2014b}, but later verified with floating-element force balance measurements}. 
\textcolor{black}{The spanwise-averaged hub-height incoming velocity $U_0$ is 1.42 ms$^{-1}$, and the incoming turbulence intensity, $TI$, is 4.8\%. 
The turbine was positioned 10 m downstream of the tunnel inlet, as shown in figure \ref{fig:WT}(b), where the boundary layer had time to fully adjust to the surface conditions and reached a fully-developed state, and quantities are slowly varying in $x$. A summary of the main experimental parameters for both cases is presented in table \ref{tab1}. \textcolor{black}{Here, the integral timescale is evaluated by integration of the autocorrelation coefficient of the velocity fluctuations until its first zero crossing, with a robust procedure similar to that described in \cite{Smith:2018}, which involves ensemble averaging the timescale over several independent realisations of the original signal. As in \cite{Gambuzza:2021}, we used signals with a duration of 200 times the initially estimated integral timescale}. Then, the non-dimensional spanwise-averaged integral lengthscale at hub height ($\Lambda/D$ in table \ref{tab1}) is obtained by applying Taylor's hypothesis of `frozen turbulence'.} 

\begin{table}
\centering
\begin{tabular}{lcccccccccc}
Case ID &Inflow &Turbine model     &D &$\left(\frac{u_*}{U_{\infty}}\right)^2\times 10^{3}$	&$z_0$		&$Re_\delta \times10^{-3}$ &$TI (\%)$ &$\Lambda/D$\\[3pt]  
(FS) Exp   &Empty tunnel &Porous disk  &416 &--	&--	 &-- &1.98&0.044\\
(BL) Exp    &Spires \& roughness &3-blade WT &416  &2.2	&0.18	&59&4.80&0.648\\
\end{tabular}
\caption{\black Summary of the experimental conditions in the reference cases (i.e., with no turbine/disk). Boundary layer characteristics are derived at $x/D=2$, \textcolor{black}{though slowly varying in $x$}. $u_*$ is the friction velocity, $z_0$, and $D$ are the roughness length and the turbine/disk diameter in mm. $Re_\delta$ is the boundary layer thickness Reynolds number, $\Lambda$ the integral lengthscale, and $TI$ is the turbulence intensity.}
\label{tab1}
\end{table}

\textcolor{black}{For both cases,} three-component velocity measurements were acquired with a 3D LDA (Dantec Dynamics, Denmark) \textcolor{black}{at a minimum frequency of 200 Hz for 120 s for each measurement point, which balanced competing requirements between data statistical convergence, reasonable running time, seeding density, and temporal resolution required to resolve the turbulence scales of interest, as further discussed in \cite{Placidi:2023b}. The three components are acquired independently, but can be synchronised by interpolation, when required by the analysis}. 
Standard errors are within $\pm 0.5\%$ and $\pm 5\%$ for the mean and second-order quantities (95\% confidence level). Three laser beams 
emanating from two laser probes (of a focal length of $300$ mm) were used in conjunction to measure the velocity components as shown in figure \ref{fig:WT}(b). One probe measured streamwise and lateral components, while the other measured the vertical component independently. A 45$^{\circ}$ mirror is used to focus the 
beam measuring the vertical component onto the same measurement volume of the other two beams, allowing for simultaneous measurements of all three velocity components. Both probes were mounted vertically in the tunnel (hence the need for the mirror) and embedded into an aerodynamic shroud to minimise flow interference. This setup helps minimise the intrusiveness of the measurement system while circumventing the error propagation originated by the 3D transformation matrix and accurate determination of the position/separation between the beams. 
Measurements, \textcolor{black}{for all cases,} were collected at different streamwise locations ($2\le x/D\le15$) both with and without the turbine\textcolor{black}{/disk model to isolate the wake-added quantities from their counterpart in the background flow.} 

\begin{figure}
    \centering
    \begin{overpic}[width=.9\textwidth]{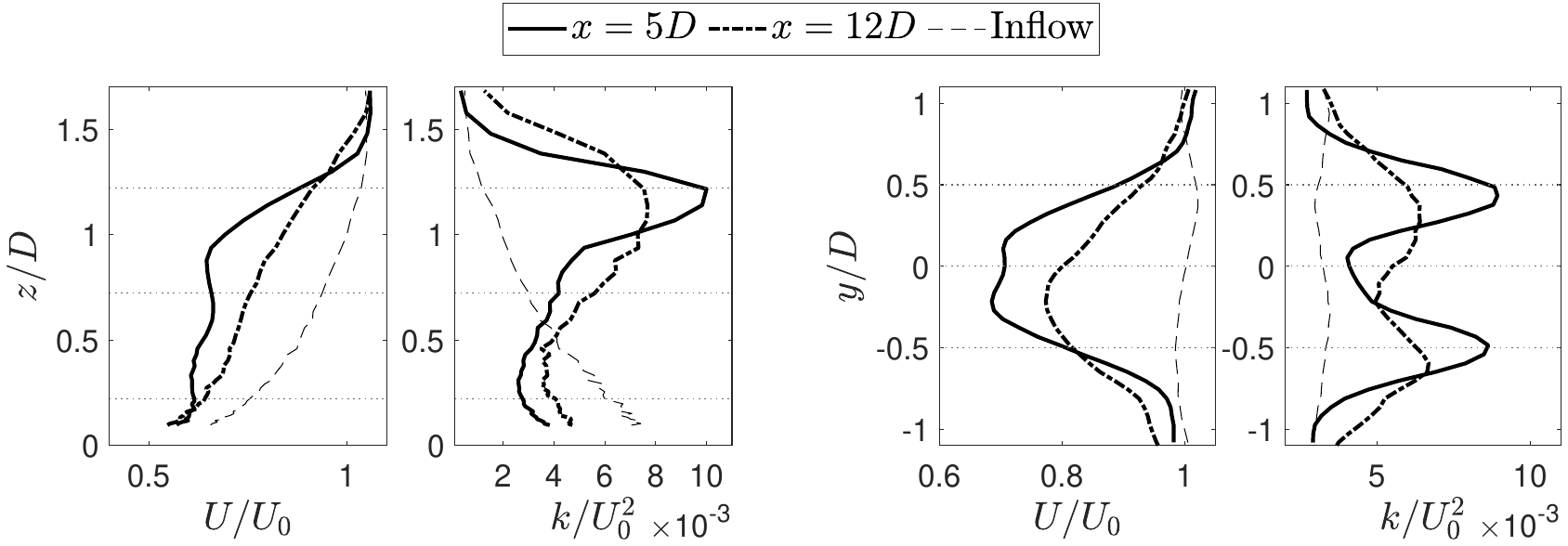}
    \put(-3,27){(a)}
    \put(52,27){(b)}
    \end{overpic}
    \caption{(a) Vertical and (b) lateral profiles of normalised streamwise velocity ($U/U_0$) and normalised TKE ($k/U_0^2$) \black for the boundary-layer case\black. Horizontal lines represent the hub and the tip positions.} 
    \label{fig:vertical_horizontal}
\end{figure}

\black Before any results are presented, we discuss the relevance of the boundary-layer case to the model developed for axisymmetric wake flows in \S\,\ref{sec:model}\black. The TKE transport equation (in \S\,\ref{sec:model}) is simplified by assuming wake flow axisymmetry and inflow homogeneity. While these assumptions are valid for the porous disk in a uniform flow, \black they do not hold in the boundary-layer case. \black Figure \ref{fig:vertical_horizontal}(a) shows vertical profiles of normalised streamwise velocity ($U/U_0$) and TKE ($k/U_0^2$) at two downwind locations \black for the boundary-layer case, \black where $z$ is the height from the ground in the Cartesian coordinate system. Inflow conditions are also reported for comparison (dashed lines). Figure  \ref{fig:vertical_horizontal}(a) shows that due to the mean shear in the incoming boundary layer, neither the inflow velocity nor the inflow TKE is uniform in the vertical direction. Looking at $x=5D$, it is evident that the wake increases the flow shear in the upper half of the wake while decreasing it in the lower half. This generates/suppresses turbulence in above/below hub height, as shown in figure \ref{fig:vertical_horizontal}(a). \black The assumptions made in \S\,\ref{sec:model} are therefore \black clearly violated in the vertical direction, and the developed model is not expected to provide satisfactory predictions. In the lateral direction, $y$, however, the incoming flow is approximately uniform, and the TKE lateral profiles appear to be more symmetrical as seen in figure \ref{fig:vertical_horizontal}(b). Therefore, despite the fact that the wake is not axisymmetric in this case, it is still of interest to examine whether the developed model can be employed to estimate lateral TKE profiles at the turbine's hub height. \black  It is also worth noting that the slight lateral wake deflection to the right as seen in figure \ref{fig:vertical_horizontal}(b) is due to the interaction of the wake swirling in the anticlockwise direction (seen from upstream) with the incoming \black flow shear \black \citep{fleming2014evaluating}.  Since equations in \S\,\ref{sec:model} are written in cylindrical coordinates, hereafter, quantities for the boundary-layer case are averaged on both sides of the wake for presentation purposes (i.e. $f(r)=0.5(f(y)+f(-y))$ for $r=y$). 

\section{Results and Discussions}\label{sec:Res}

\begin{figure}
    \begin{center} 
        \includegraphics[width=.9\linewidth]{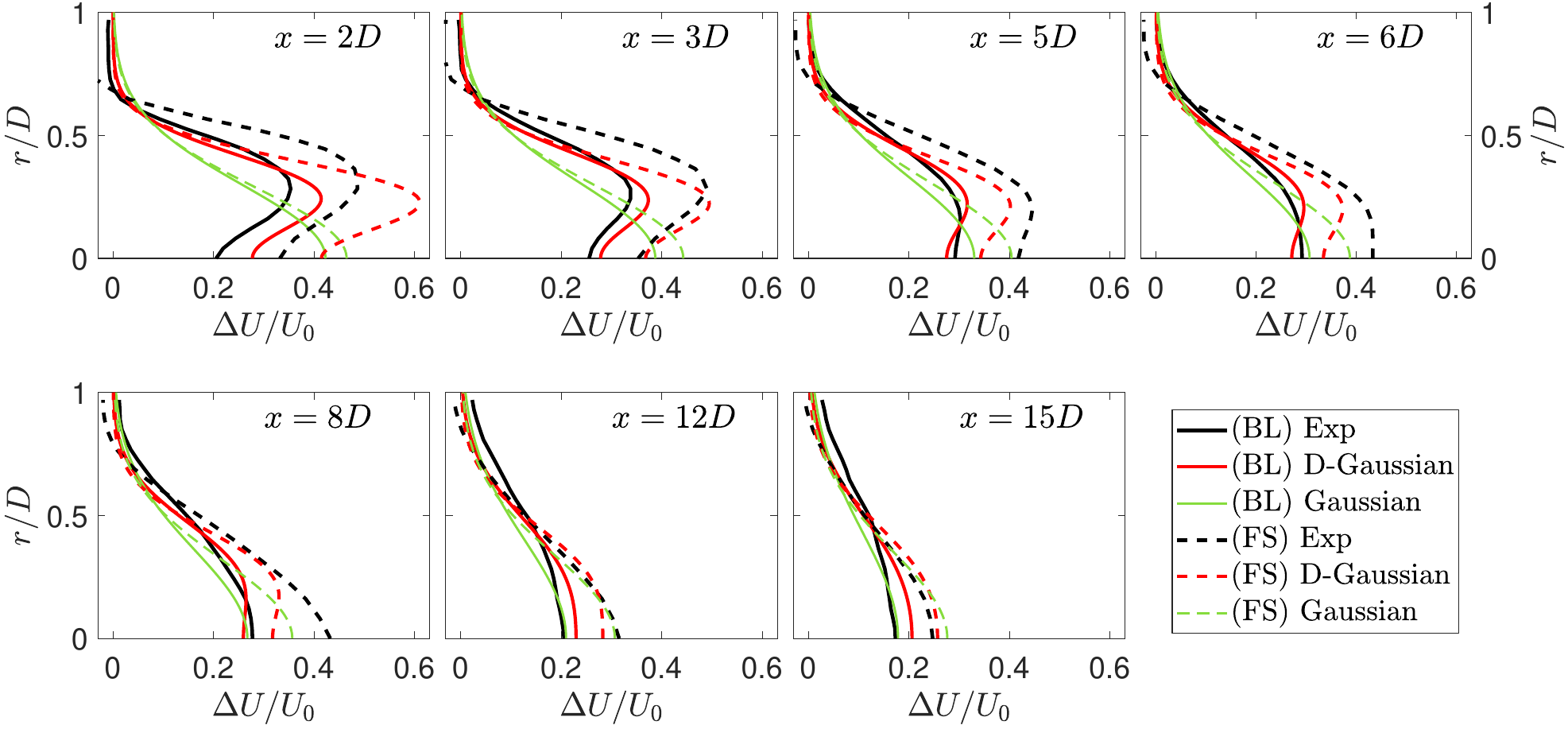}
        \captionof{figure}{Radial profiles of normalised velocity deficit ($\Delta U/U_0$) at different $x/D$ \black based on experiments, Gaussian and double-Gaussian (referred to as D-Gaussian) wake models. \black Solid lines show the boundary-layer (BL) case and dashed lines show the freestream (FS) case. }
        \label{fig:DeltaU}
    \end{center}
\end{figure}

To predict TKE profiles based on \eqref{eq:modsol}, \black the wake velocity profile $U(x,r)$ needs to be first known either from experiments or engineering wake models. Figure \ref{fig:DeltaU} shows radial profiles of the normalised streamwise velocity deficit, $\Delta U/U_0=(U-U_0)/U_0$ at different streamwise locations \black for both cases \black and compares them to two customary profiles: Gaussian \eqref{gauss} and double-Gaussian \eqref{doublegauss}. 
\black The figure shows that the velocity deficit in the boundary-layer case is consistently smaller than the other case. 
The observed difference is expected to be caused by the different inflow conditions \red and thrust coefficients \black in the two cases. A higher level of inflow turbulence \red and a lower value of thrust coefficient \black in the boundary-layer case leads to a \red less pronounced \black wake. \black To predict wake velocity profiles, the double-Gaussian profile \eqref{doublegauss} is used hereafter as \black it generally better represents our experimental data for $x/D \leq 6$ where the wake \black profiles \black present a double-peak, \black although some discrepancies are noticeable especially for the free-stream case.  Further work can be done to see if the the double-Gaussian model parameters can be better optimised to improve its predictions. \black It is important to note that inaccuracies in velocity prediction in the near wake manifest as an error in TKE predictions for the far wake. This highlights the importance of accurate velocity predictions in the near wake. Depending on the actual shape of the near-wake profiles, one may also apply other wake models such as super-Gaussian \citep{shapiro2019paradigm,blondel2020}. This is possible due to the versatility of the new model developed for a generic profile of $U(x,r)$.

\begin{figure}
    \centering
    \begin{overpic}[width=\linewidth]{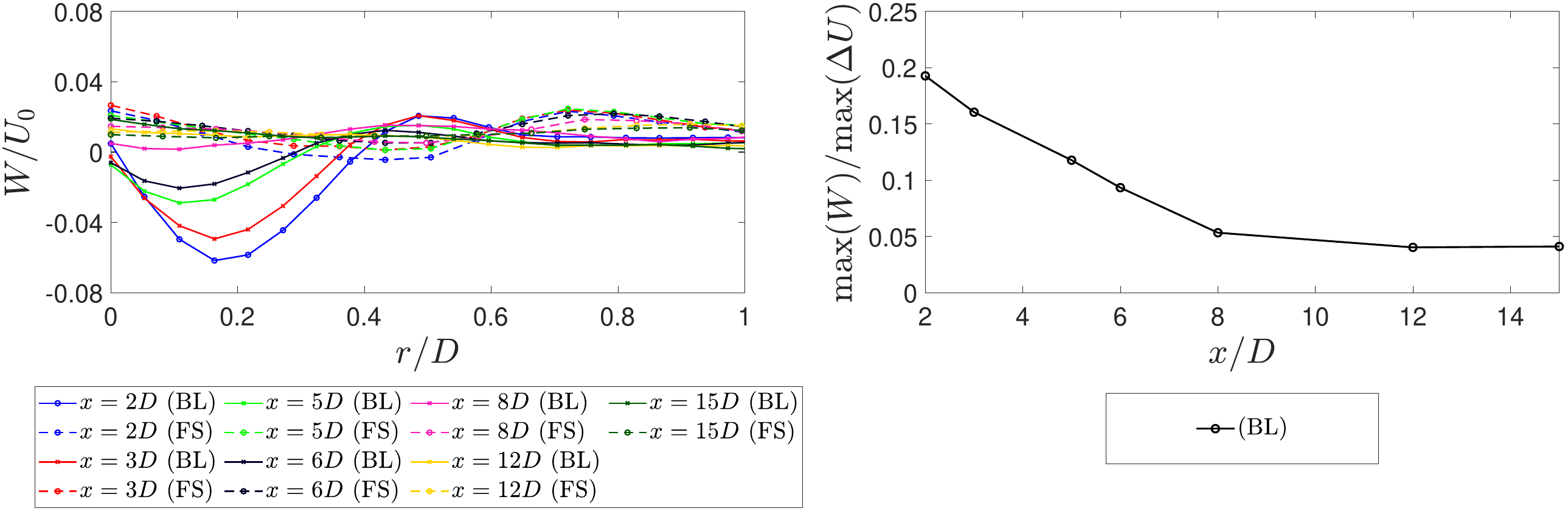}
        \put(-,31){(a)}
        \put(49,31){(b)}
         \end{overpic}
        \caption{(a) Lateral profiles of normalised azimuthal velocity ($W/U_0$) at hub height. (b) Ratio of the maximum $W$ to the maximum $\Delta U$ at different $x/D$. \black BL denotes the boundary-layer case, and FS is the freestream case.}
    \label{fig:W}
\end{figure}

\black Next, the radial profiles of the normalised azimuthal velocity $W/U_0$  at different streamwise positions is shown in figure \ref{fig:W}(a) for both cases. The azimuthal velocity in the freestream case is negligible as the porous disk does not generate any swirl motion. In the boundary-layer case, while rotating blades of the turbine induce swirl motion in the wake, the value of wake swirl decays rapidly. Figure \ref{fig:W}(b) shows that the maximum azimuthal velocity in the wake is considerably smaller than the maximum velocity deficit, especially in the far-wake region. This is consistent with the TKE budget analysis, discussed later, which will show that all terms in the TKE budget \eqref{eq:TKE-transport1} that include azimuthal velocity have negligible values at $x=5D$. \black Nonetheless, it is still important to bear in mind that despite the rapid decay of swirl and seemingly its small contribution in the far wake, the wake swirl in the near wake may still have a non-negligible impact on the TKE in the far wake.\black


\begin{figure}
    \centering
    \includegraphics[width=.9\textwidth]{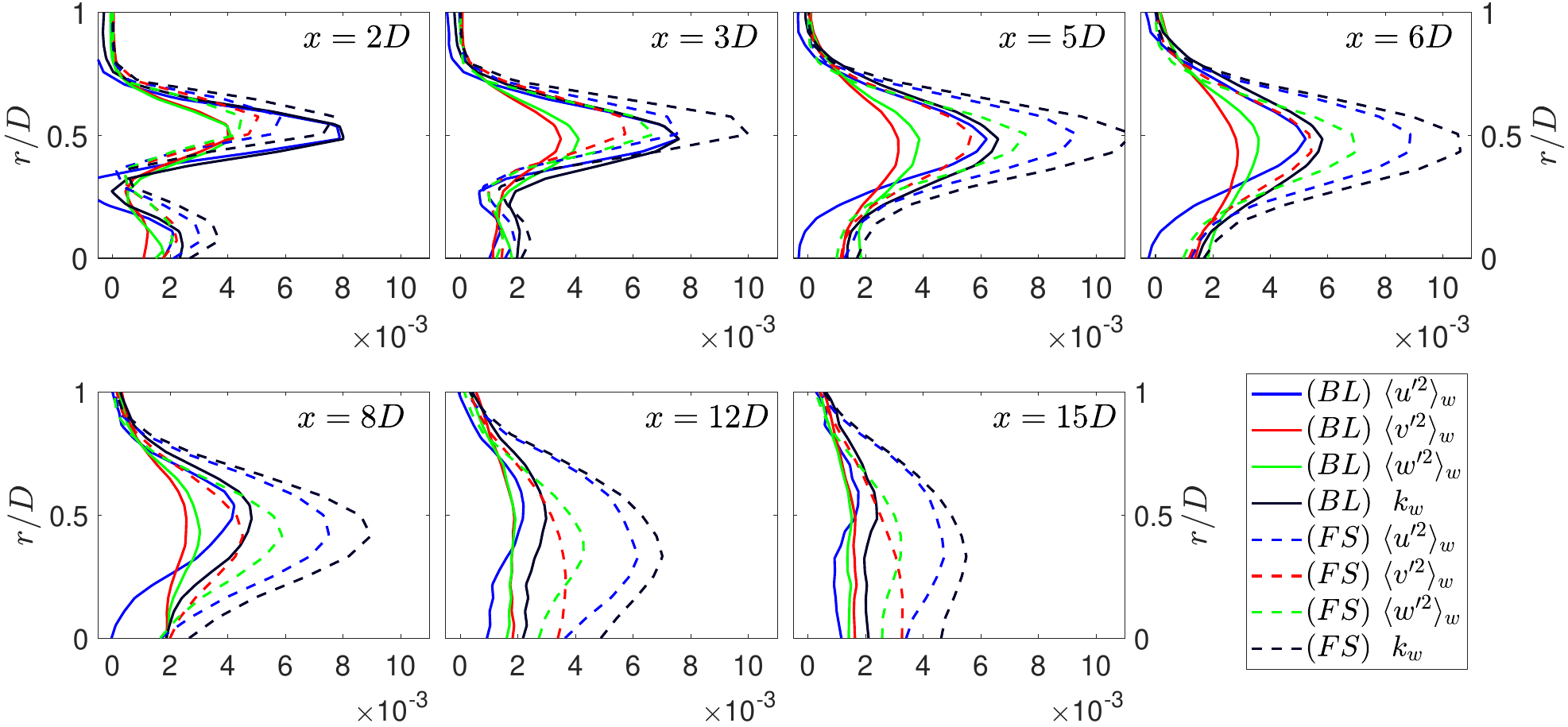}
    \caption{Radial profiles of normalised wake-added turbulent 
quantities at different $x/D$. \black Solid lines show the boundary-layer (BL) case and dashed lines show the freestream (FS) case. 
}
    \label{fig:flucts}
\end{figure}

Figure \ref{fig:flucts} shows radial profiles of the wake-added turbulent velocity fluctuations ($\langle u^2 \rangle_w$, $\langle v^2 \rangle_w$, and $\langle w^2 \rangle_w$) and the wake-added TKE ($k_w$) at different streamwise locations \black for both cases. \black 
It is clear how the wake edge (i.e. $r\approx 0.5 D$) \black where the shear production is maximum \black corresponds to the maximum level of turbulence. \black At the wake edge, $\langle u^2 \rangle_w$ is dominant and significantly larger than $\langle v^2 \rangle_w$ and $\langle w^2 \rangle_w$ in both cases. \black  This is especially the case in the near wake of the boundary-layer case.  However, the other two components ($\langle v^2 \rangle_w$, and $\langle w^2 \rangle_w$) gradually become larger and more comparable to $\langle u^2 \rangle_w$ in the far wake. \black \cite{spalart1988direct} showed that, for boundary-layer flows, the TKE diffusion due to fluctuating pressure is mainly small, however, pressure has a significant role in redistributing the energy by extracting it from the streamwise component and transferring it to the other two components \citep{pope2000}. This seems to be the case here too. It is also interesting to note that, at the wake centre ($r/D<0.25$) of the boundary-layer case, 
$\langle v^2 \rangle_w$ and $\langle w^2 \rangle_w$ are bigger than $\langle u^2 \rangle_w$ for $x>3D$. Cross-stream turbulent fluctuations are especially critical in explaining unsteady oscillations of wakes termed as wake meandering \citep{larsen2008}, \black which has major impacts on flow mixing and wake expansion. \black However, prior experiments have often quantified the TKE distribution only based on streamwise fluctuations. Moreover, steady-state engineering wake models have often used the turbulence intensity defined only based on streamwise fluctuations to estimate the wake expansion. \black This highlights the importance of the current experimental work in capturing the total TKE. 


Next, we determine three important turbulent quantities using the experimental data: (i) turbulent viscosity ($\nu_t$), (ii) mixing length ($l_m$) and (iii) the normalised TKE dissipation rate ($C_\varepsilon$). The last two are needed to compute the parameter $\Psi=cl_m^2/C_\varepsilon$ in the simplified TKE transport equation \eqref{eq:TKE-transport2}. 
Both turbulent viscosity and mixing length are estimated based on the method described in \cite{Bai:2012} and later implemented in \cite{Rockel:2016} and \cite{Scott:2023}. At each streamwise location, $\nu_t$ is the slope of the linear curve fitted to the variation of $-\langle uv\rangle$ with respect to $\partial U/\partial r$, and $l_m^2$ is the slope of the linear curve fitted to the variation of $-\langle uv\rangle$ with respect to $|\partial U/\partial r|\partial U/\partial r$. 

\begin{figure}
    \centering
   \begin{overpic}[width=.85\linewidth]{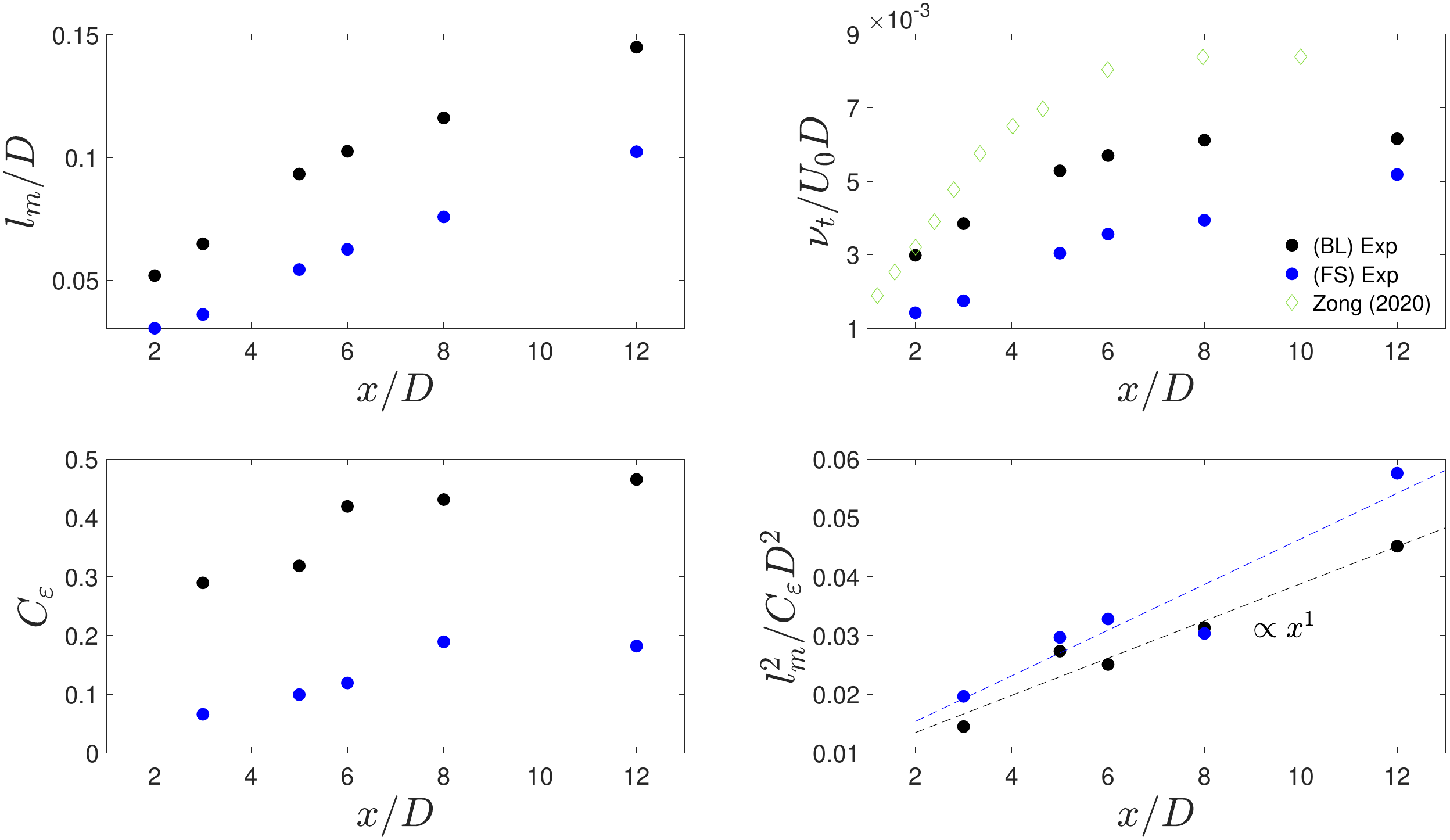}
       \put(-2,55){(a)}
       \put(50,55){(b)}
       \put(-2,25){(c)}
       \put(50,25){(d)}
   \end{overpic}
    \caption{\black Evolution of normalised (a) mixing length ($l_m/D$), (b) turbulent viscosity ($\nu_t/U_0D$) including data reported by \citet{zong2020}, (c) energy dissipation rate ($C_\varepsilon$), and (d) $l_m^2/C_\varepsilon D^2$, with $x/D$. The black circles are from the boundary-layer (BL) case, and the blue circles are from the freestream case (FS).}
    \label{fig:turbulent_viscosity}
\end{figure}
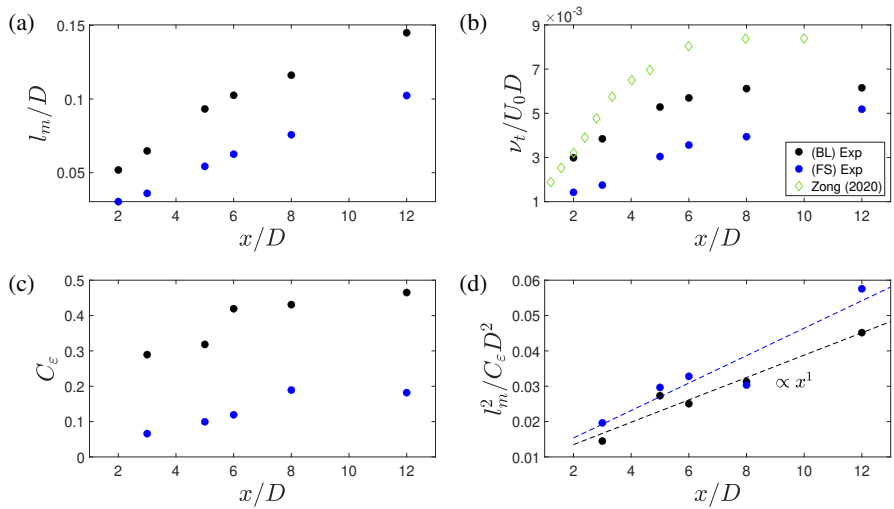

Figure \ref{fig:turbulent_viscosity}(a) shows that, \black in both cases \black as expected, the normalised mixing length increases with $x/D$, which indicates an increased characteristic length scale as the wake flow evolves \citep{Iungo:2015}. \black The trend is almost identical for the freestream case, but with a lower value at each $x/D$, which would be expected because of the lower value of inflow integral lengthscale in this case (see table \ref{tab1}). \black The normalised turbulent viscosity, shown in \textcolor{black}{figure} \ref{fig:turbulent_viscosity}(b), also \black linearly increases with $x$ and seems to approach an almost constant value in the far wake. The same trend is shown in experiments by \citet{zong2020}, displayed in green. 
In all three cases shown in figure \ref{fig:turbulent_viscosity}(b), the turbulent viscosity reaches a plateau at about $6D-8D$ downstream. 
Experiments and numerical simulations by \citet{Scott:2023} (not shown here) also reported a fairly similar behaviour, and found that further downstream (e.g. $x/D>>15$), the turbulent viscosity should decrease as the wake recovers. \black This behaviour of $\nu_t$ can be explained with the hypothesis that $\nu_t \approx l_m^2|\partial U/\partial r|$  \citep{Bai:2012}. In the near wake, the increase in the turbulent viscosity is mainly due to the increase in the mixing length. However, in the far wake, the growth in the mixing length is balanced by the decreasing velocity gradients across the wake. \black It is also noteworthy that \red despite having higher $C_T$\black, the turbulent viscosity is smaller in the free-stream case compared with the boundary-layer case, which can be attributed to a lower level of inflow turbulence in the former case. On other hand, the turbulent viscosity in \cite{zong2020} is higher than the boundary-layer case although both have a fairly similar inflow turbulence level. The discrepancy between these two cases can be explained by the \red different \black thrust coefficients of the turbine models. In conclusion, figure \ref{fig:turbulent_viscosity}(b) shows that while the variation of $\nu_t$ with $x$ follows a fairly similar pattern in all cases, its value \red is sensitive to \black both inflow turbulence and thrust force.\black 

\begin{figure}
    \centering
   \includegraphics[width=\textwidth]{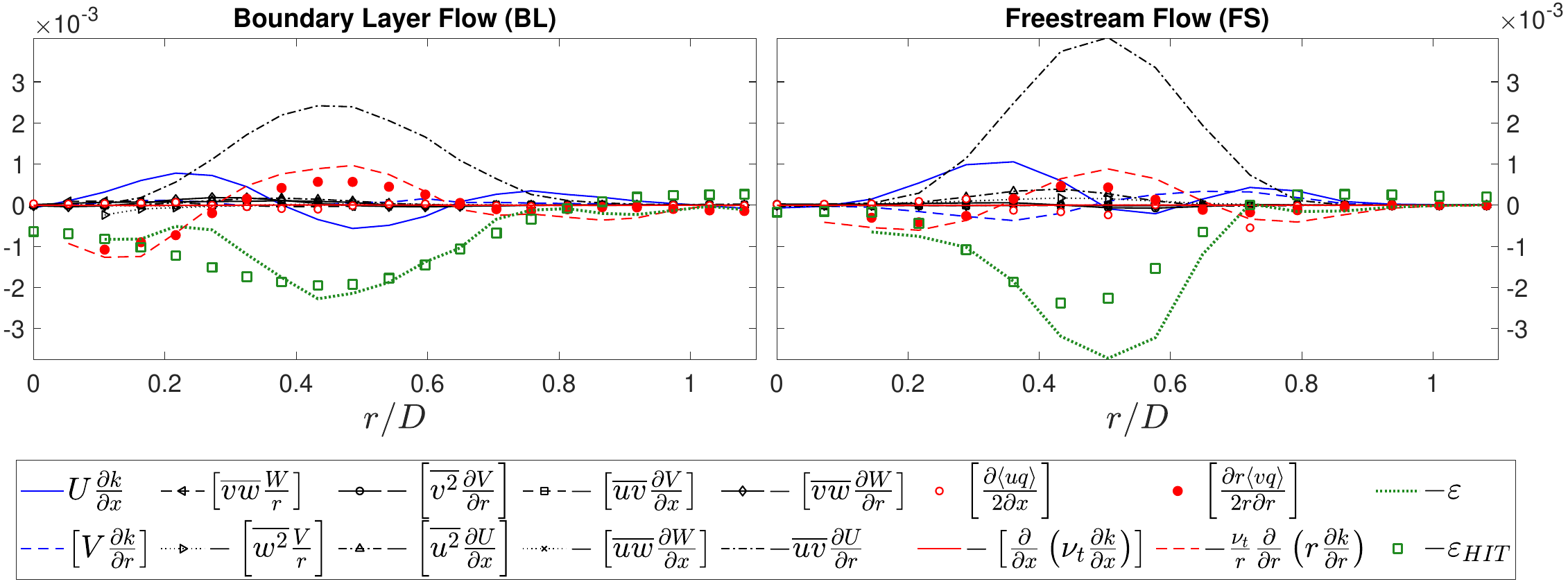}
    \caption{TKE budget for both cases at $x/D=5$ (terms normalised by $U_0^3/D$). Advection (blue), diffusion (red), production (black), dissipation (green).}
    \label{fig:TKe_budget}
\end{figure}

Next, we compute the TKE dissipation rate ($\varepsilon$, hereafter \textit{dissipation} in short). Estimating the dissipation from experimental data is a troublesome task \citep{Wang:2021}. Two common approaches are used here. In the first method, thanks to our comprehensive experimental data set, we estimate the dissipation \textit{indirectly} by performing a TKE budget analysis, i.e. evaluating all the terms in \eqref{eq:TKE-transport1}, so that by definition its residual is the dissipation \citep{Hearst:2014}. 
As an example, the TKE budget is provided in figure \ref{fig:TKe_budget} for $x=5D$. 
Figure \ref{fig:TKe_budget} demonstrates that all the terms in square brackets in \eqref{eq:TKE-transport1} can, indeed, be considered negligible \black for both cases. The diffusion terms are shown in red, where the real terms are represented as circles and the modelled terms based on the gradient-diffusion hypothesis in \eqref{eq:gradient-diffusion} as lines. It can be seen that in both cases, the radial diffusion term, which is the dominant diffusion mechanism in the wake and its modelled counterpart behave similarly. 
\black 
 The figure also suggests that the streamwise convection term is in balance with the radial diffusion term at this streamwise location, \black which is particularly evident in the boundary-layer case. \black The radial diffusion term is positive at the wake edge, acting as a sink of energy transporting energy towards the wake's centre and the outer region where the radial diffusion is negative (i.e. acting as an energy source). This process flattens and widens the TKE profiles as the wake moves downstream, as already seen in figure \ref{fig:flucts}. The dominant production term (dot-dashed black line) is mainly in balance with the dissipation (green dotted line) as the main two dominant terms and, as expected, $-\varepsilon$  is negative so that it acts as an energy sink. Both have maximum absolute values in correspondence of the blade tip at the wake edge ($r/D\approx 0.5$).

Given the time-resolved nature of the experimental data, we can also \textit{directly} quantify the dissipation based on another common approach that assumes the turbulence at small scales to be homogeneous and isotropic (subscript $_{HIT}$). This leads to 
the estimation of the dissipation based on Taylor's hypothesis of frozen turbulence, as $\varepsilon_{HIT}=15\nu\langle(\partial u/\partial t)^2\rangle/U^2 $, where $\nu$ is the kinematic viscosity and $t$ is time \citep{Dairay:2015,neunaber2022application}. Figure \ref{fig:TKe_budget} shows there is a \textcolor{black}{satisfactory} agreement between the two methods used herein to estimate the dissipation, \black particularly in the boundary-layer case. Although there is a difference in the magnitude between the two dissipation terms at the wake edge ($r/D\approx 0.5$) for the freestream case, overall there is a satisfactory agreement given a high level of uncertainty in estimating the dissipation. \black

Once the dissipation is estimated, the normalised TKE dissipation rate ($C_\epsilon$) can be computed based on $\epsilon l_m/K_w^{3/2}$, where $\epsilon$ and $K_w$ are the maximum absolute values of $\varepsilon$ and $k_w$ at each streamwise location, respectively. Figure \ref{fig:turbulent_viscosity}(c) confirms that $C_\varepsilon$ is, indeed, not constant and instead increases with $x$. 
This is in line with previous works that highlighted how $C_\varepsilon$ may vary in non-equilibrium flows \citep{obligado2016nonequilibrium,Dairay:2015,vassilicos2015review}. \black Moreover, in line with other turbulent quantities, the normalised TKE dissipation rate is higher for the boundary-layer case. \black Next, figure \ref{fig:turbulent_viscosity}(d) reports the variation of $l_m^2/C_\varepsilon D^2$ with $x/D$. Based on the fitted linear curve shown \black by the black dashed line \black in the figure, $l_m^2/C_\varepsilon D^2\approx 0.0072+0.0032x/D$ \black for the boundary-layer case \black. As discussed in \S\,\ref{sec:model}, $\Psi=cl_m^2/C_\epsilon$, so $\Psi_{BL}/D^2\approx c(0.0072+0.0032x/D)$. \black Similarly, for the freestream case (blue dashed line), $\Psi_{FS}/D^2\approx c(0.0076+0.0039x/D)$, \black where $c=0.46$ seems to provide satisfactory predictions \black for both. \black It is interesting to note that this value is comparable to $c=0.55$ used in the log layer of boundary layer flows  \citep{pope2000}. \black Furthermore, despite significant disparities in other turbulent parameters such as $l_m$ and $C_{\varepsilon}$, the value of $\Psi$ appears fairly similar in both cases. This similarity tempts us to speculate about the existence of a universal relationship for $\Psi$. However, further investigation is required to substantiate this hypothesis.\black 

\begin{figure}
    \centering
    \includegraphics[width=.85\textwidth]{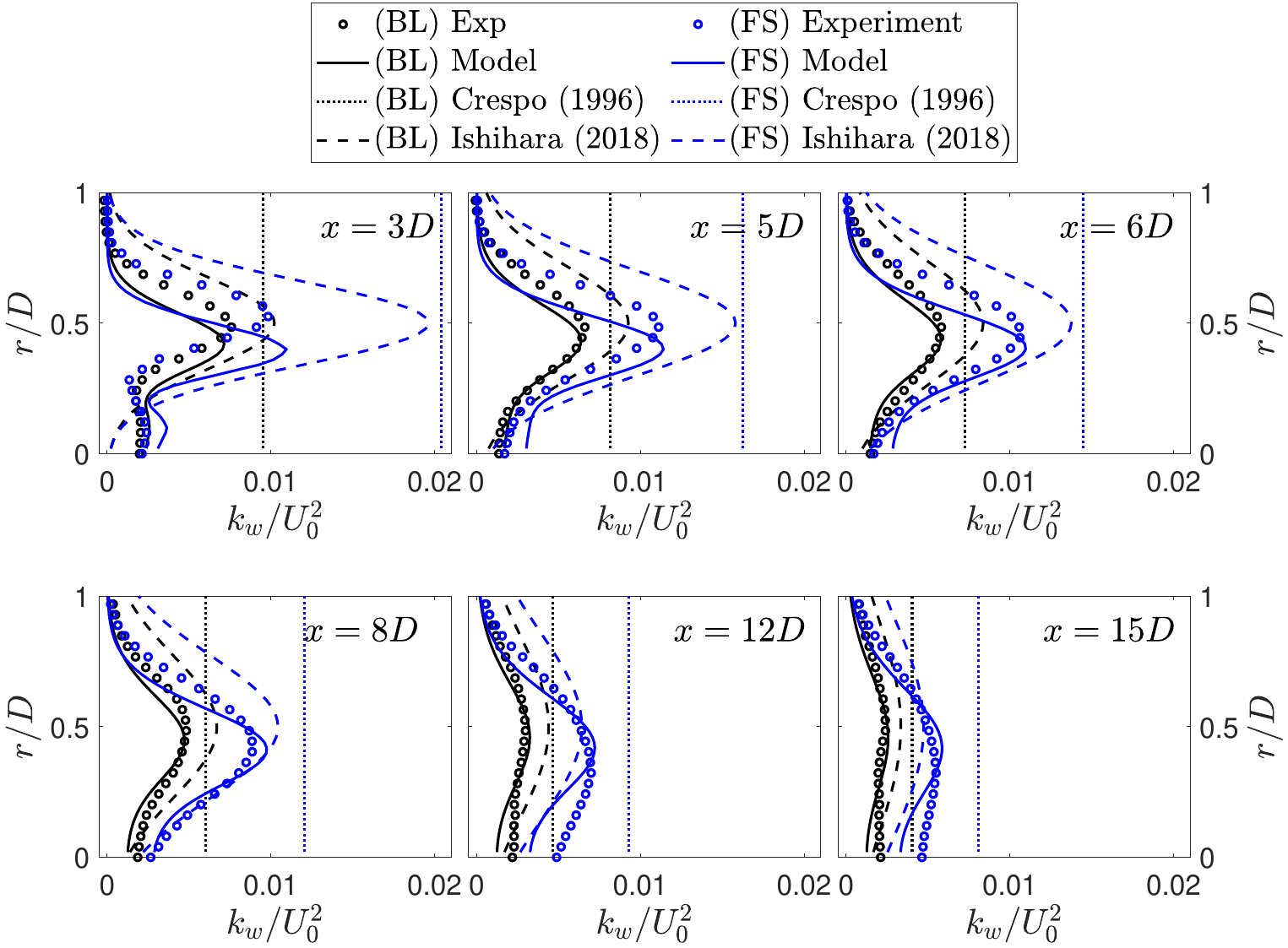}
    \caption{Model predictions of radial profiles of normalised wake-added TKE ($k_w/U_0^2$) in comparison with the experimental data \black and the empirical models of \cite{crespo1996} and \cite{ishihara2018new} at different streamwise positions. The boundary-layer (BL) case is shown in black, and the freestream case (FS) in blue. 
}
    \label{fig:TKe_model}
\end{figure}

\black After assessing both $\nu_t(x)$ and $\Psi(x)$, we proceed to compare the predictions of the model outlined in \S\,\ref{sec:model} with the experimental data, depicted in figure \ref{fig:TKe_model}. Some deviations are apparent in predicting the location of TKE maxima in the near wake for the freestream case and also the TKE level at the wake centre in the far wake for both cases. These discrepancies are believed to primarily stem from inaccuracies in the velocity-deficit predictions illustrated in figure \ref{fig:DeltaU}. Nevertheless, the overall trend reveals that the model reasonably predicts both the magnitude of wake-added TKE ($k_w$) and its radial distribution for $3\leq x/D\leq 15$ in both cases. Moreover, the model successfully captures the \red difference 
 in \black wake-added TKE \red between the two cases. \black Figure \ref{fig:TKe_model} shows that the wake-added TKE is considerably higher in the freestream case compared to the boundary-layer case. As previously discussed, our results show that the inflow turbulence \red and the amount of the thrust force \black significantly impact all pertinent turbulent quantities. 
  Thus, predicting \red their \black overall effects on wake-added TKE is not straightforward, given their counteracting influences on wake behaviour. This emphasises the necessity of employing a physics-based TKE model capable of realistically predicting wake behaviour under varied \red operating \black conditions. It is worth highlighting that while the freestream case shows a $3\%$ lower turbulence intensity compared to the boundary-layer case (as indicated in table \ref{tab1}), the difference becomes significant when considering the integral length scale, where there is more than a one order of magnitude difference. This emphasises that relying solely on turbulence intensity \red may \black not fully capture the impact of the incoming turbulent flow on the wake evolution \citep{Gambuzza:2021,hodgson2023effects}.

Predictions based on the empirical models of \cite{crespo1996} and \cite{ishihara2018new} are also shown in figure \ref{fig:TKe_model}. Since these two models predict the wake-added turbulence only based on streamwise velocity fluctuations, for a fair comparison, we use the relationship $\sqrt{\langle u_w^2\rangle}=\alpha \sqrt{k_w}$, where $\alpha$ is a constant \citep{larsen2022calculating}. Various values for $\alpha$ have been suggested in the range $\alpha \in [0.82 \ 1.03]$ \citep[e.g.][]{crespo1996, malki2014planning,Cleijne1992ResultsOS}. Our experimental data suggests that $\alpha \approx 0.93$ and is used to plot the total wake-added TKE ($k_w$) predictions for these two empirical models in figure \ref{fig:TKe_model}. The predictions based on the model of \cite{crespo1996} only depends on the streamwise distance from disk/turbine, while the one proposed by \cite{ishihara2018new} also predicts the radial distribution of wake-added turbulence. Both empirical models 
cannot \red fully \black capture the impact \red of changing operating conditions \black on the wake-added TKE distribution.  \black

\section{Summary}\label{sec:summary}
A new fast-running model to predict the 3D TKE distribution in axisymmetric wake flows is presented. \black Detailed 3D LDA measurements were conducted for two canonical cases: (i) a porous disk exposed to a uniform freestream flow, and (ii) a turbine model under a turbulent boundary layer. While the former configuration generates an axisymmetric wake, the wake flow in the latter case is non-axisymmetric due to inflow shear. In the latter case, our analysis primarily focused on the flow distribution within a horizontal plane at hub-height level. \black A budget analysis is first performed to identify dominant terms in the TKE transport equation written in a cylindrical coordinate system. The Boussinesq turbulent-viscosity and gradient-diffusion hypotheses were used to simplify production and diffusion terms, respectively. The simplified partial differential equation was then solved using the Green function's method, which led to a solution written in the form of a double integral. Further simplifications were applied to the exact mathematical solution to facilitate the numerical integration. \black The new model requires numerical integration based on simple methods such as the trapezoid rule, which can be performed with very basic mathematical knowledge. Therefore, in addition to addressing computational costs, the ease of use (in comparison with CFD models) is the main driving factor behind the development of such a model. \black  The developed solution predicts second-order flow statistics (i.e. TKE distribution) from the knowledge of first-order flow statistics (i.e. time-averaged streamwise velocity distribution, $u(x,r)$). While the solution was derived for an arbitrary distribution of $u(x,r)$, a double-Gaussian profile was assumed herein due to its resemblance to the experimental data, especially in the near wake. To predict the TKE, the model also necessitates the turbulent viscosity $\nu_t(x)$, and the parameter $\Psi(x)=cl_m^2/C_\varepsilon$. 

The experimental data showed that \black in both cases \black the mixing length in the wake flow grows with the streamwise distance from the disk/turbine. The increase in the mixing length initially leads to an increase in turbulent viscosity, but the turbulent viscosity approaches a constant value in the far wake as wake velocity gradients diminish with the wake recovery.  \black \red Operating conditions (i.e. $C_T$ and \black the inflow turbulence) are found to have major impacts on turbulent quantities such as mixing length and turbulence viscosity. \black Moreover, in agreement with non-equilibrium similarity theory \citep{vassilicos2015review}, we observed that the normalised TKE dissipation rate ($C_\varepsilon$) is indeed not constant, but it increases with streamwise distance from the turbine; this results in a linear relationship for $\Psi(x)$. Finally, the new TKE model predictions are compared to the experimental data demonstrating a \black satisfactory \black level of agreement both in the magnitude and the radial shape of the TKE profiles across a wake flow extent of interest to wind farms developers and operators ($3<x/D<15$). 

\black This section is concluded by a discussion on model limitations and future research directions. In this work, we relied on the same experimental data to determine the variations of $\nu_t(x)$ and $\Psi(x)$ with $x$, which were then used to validate model predictions. This validation does not inherently establish the universality of the relationships governing these input parameters. Therefore, we opt not to claim that the same parameter settings are universally applicable across all different scenarios. Our main goal in this study was to demonstrate the robustness of the developed model. Specifically, we aimed to show that when accurately estimating these parameters, our model reliably predicts the distribution of TKE in axisymmetric wake flows (and approximates non-axisymmetric turbine wake flows at the hub-height level). However, we acknowledge the need for further research to establish universal relationships for $\nu_t$ and $\Psi$ across a range of relevant parameters, thereby creating a comprehensive framework for TKE engineering modeling.

The developed model uses information on mean streamwise velocity distribution as an input to estimate TKE generation resulting from flow shear. Integrating this model into existing engineering velocity-deficit models that determine the wake recovery rate based on the incoming turbulence is straightforward. However, this approach does not take into account the two-way coupling effect, where velocity gradients in the wake generate higher turbulence levels, which subsequently influence mean flow distribution through enhanced flow mixing. Investigating how the wake-added TKE may impact flow entrainment and wake recovery presents an interesting area of research \citep{nygaard2020modelling}. Future studies could potentially explore this by using the developed model in an iterative approach that considers the interplay between first-order and second-order statistics. Alternatively, addressing this coupling in the developed TKE model may involve further simplification and decoupling of equations, leveraging assumptions such as the self-similarity of TKE profiles. In addition, more refined models are needed to predict vertical TKE profiles in non-axisymmetric wakes of turbines immersed in boundary-layer flows. \black 

\backsection[Acknowledgements]{We would like to thank Dr Paul Hayden, who aided the experimental data collection.}
\backsection[Funding]{The experimental work was supported via Flex Funding agreement project VENTI within EPSRC-Supergen ORE (EP/S000747/1) coordinated by the University of Plymouth.}
\backsection[Declaration of interests] {The authors report no conflict of interest.}
\backsection[Data availability statement]{\textcolor{black}{The wind tunnel data utilised in this work is available at the following link: https://DOI: 10.15126/surreydata.901120.}}
\appendix

\section{\black Green's function solution of the TKE transport equation}\label{appA}\black
The derivation of \eqref{eq:modsol} is as follows. {\black{First, we use an integrating factor to cancel the $k_w(x,r)$ term (i.e. the third term on the left-hand side) in \eqref{eq:TKE-transport2}. Set}} $k_w(x,r)=\exp\{-\psi(x_0,x)\}Q(x,r)$,
which reduces \eqref{eq:TKE-transport2} to
\begin{equation}\label{TKEmod1}
 		\frac{U_0}{\nu_t(x)}\frac{\partial Q(x,r)}{\partial x}-\frac{1}{r}\frac{\partial}{\partial r}\left( r\frac{\partial Q(x,r)}{\partial r} \right) = \bigg(\frac{\partial U({\black{x}},r)}{\partial r}\bigg)^2\exp\{\psi(x_0,x)\}.	
 \end{equation}
{\black{Now introduce}} a change of variable $T(x)=\phi(x_0,x)$ and a corresponding dummy variable $t(X)=\phi(x_0,X)$ as a coordinate on the interval $0\leq t\leq T$. As $\nu_t>0$ throughout the region of interest, $t$ is a monotonically increasing function of $X$; consequently, $X$ can be used as a coordinate in the range $x_0\leq X\leq x$.

 Let $Q(x,r)=q(T(x),r)$, which reduces \eqref{TKEmod1} to {\black{an inhomogeneous axisymmetric heat equation:}}
 \begin{equation}\label{TKEmod2}
 	\frac{\partial q(T,r)}{\partial T}-\frac{1}{r}\frac{\partial}{\partial r}\left( r\frac{\partial q(T,r)}{\partial r} \right) = \bigg(\frac{\partial U({\black{x}}(T),r)}{\partial r}\bigg)^2\exp\{\psi(x_0,x(T))\}.	
 \end{equation}
 {\black{The Green's function for such equations,}} subject to $q(0,r)=0$, $\partial q/\partial r(T,0)=0$ and $q(T,r)\rightarrow 0$ as $r\rightarrow \infty${\black{, is \citep[see][]{cole2011}
 \[
 G(T,r;t,\rho)=\frac{H(T-t)}{4\pi(T-t)}\,\exp\left\{-\,\frac{r^2+\rho^2}{4(T-t)}\right\}I_0\left(\frac{r\rho}{2(T-t)}\right),
 \]
 where $H$ is the Heaviside step function and $I_0$ is the modified Bessel function of the first kind. Integrating over the cylindrical domain yields the solution of \eqref{TKEmod2}:
 }}
 \begin{equation}\label{eq:modsol1}
 	q(T,r)=\int_{t=0}^T\int_{\rho=0}^\infty\frac{1}{2(T-t)}\,\exp\left\{-\,\frac{r^2+\rho^2}{4(T-t)}\,+\psi(x_0,X(t))\right\}I_0\left(\frac{r\rho}{2(T-t)}\right)\bigg(\frac{\partial U(X({\black{t}}),\rho)}{\partial \rho}\bigg)^2\rho\,\mathrm{d}\rho\,\mathrm{d}t{\black{.}}
 \end{equation}
 Changing the dummy variable from $t$ to $X$ yields the solution \eqref{eq:modsol} for $k_w(x,r)$.


\black

\section{Numerical integration of \eqref{eq:modsol}}\label{appB}
To ease the numerical integration of \eqref{eq:modsol} for an arbitrary wake velocity profile, we apply the changes below to the exact solution: 
\begin{itemize}
\item For large values of $\rho$, the Bessel function in \eqref{eq:modsol} goes to infinity, while the exponential term goes to zero, which may introduce errors in the numerical integration. For $r>0$, it is useful to write the solution \eqref{eq:modsol} in the following form
 \begin{multline}\label{modsol2}
k_w(x,r)=\frac{1}{U_0}\int\limits_{X=x_0}^x\int\limits_{\rho=0}^\infty\nu_t(X)e^{-\psi(X,x)}\left\{\frac{e^{-\,\frac{(r-\rho)^2}{4\phi(X,x)}}}{\sqrt{4\pi\phi(X,x)}}\right\}\\ \times\left[\sqrt{\tfrac{\pi r\rho}{\phi(X,x)}}e^{-\,\frac{r\rho}{2\phi(X,x)}}I_0\left(\tfrac{r\rho}{2\phi(X,x)}\right)\right] \bigg(\frac{\partial U(X,\rho)}{\partial \rho}\bigg)^2\,\sqrt{\tfrac{\rho}{r}}\,\,\mathrm{d}\rho\,\mathrm{d}X.
\end{multline}
The expression in square brackets is a well-behaved function, $f(z)$, where $z=r\rho/\phi(X,x)$, whose graph rises rapidly from $0$ up to around $1.2$ at $z\approx 1.5$, then decreases towards its limiting value of $1$ as $z$ increases further. By combining the power series and asymptotic expansions for the modified Bessel function, one obtains the following good approximation,
\begin{equation}\label{eq:f(z)}
	f(z)\approx\begin{cases}
		\sqrt{\pi z}\,\exp(-z/2)\,(1+\frac{z^2}{16}+\frac{z^4}{1024}),&0\leq z \leq 4;\\
		1+\frac{1}{4z}+\frac{9}{32z^2}\,,&z>4,
	\end{cases}
\end{equation}
which gives a maximum error of less than $1.3\%$. 
\item The integrand in the exact solution \eqref{eq:modsol} has a singularity at $X=x$, because $\phi(x,x)=0$. To avoid this singularity, we compute the integral by restricting $X$ to the interval $[x_0,x-\delta]$, where $\delta$ is the size of the grid used for numerical integration. This approximation provides satisfactory results with negligible error for small values of $\delta$ (e.g. $\delta\leq 0.1D$, where $D$ is the diameter of the object causing the turbulent wake).
\item In the exact solution \eqref{eq:modsol}, the upper bound of integration with respect to $\rho$ is infinity. For the numerical integration, we replace this with a large finite value, namely $3D$. The velocity gradient $\partial U(X,\rho)/\partial \rho$ quickly goes to zero for large values of $\rho$, so this has a negligible effect on final results. 
\end{itemize}

In summary, instead of the exact solution \eqref{eq:modsol}, one can numerically compute
 \begin{multline}\label{modsol3}
k_w(x,r)=\frac{1}{U_0}\int\limits_{X=x_0}^{x-\delta}\int\limits_{\rho=0}^{3D}\nu_t(X)e^{-\psi(X,x)}\left\{\frac{e^{-\,\frac{(r-\rho)^2}{4\phi(X,x)}}}{\sqrt{4\pi\phi(X,x)}}\right\}f\left(\frac{r\rho}{\phi(X,x)}\right) \bigg(\frac{\partial U(X,\rho)}{\partial \rho}\bigg)^2\,\sqrt{\tfrac{\rho}{r}}\,\,\mathrm{d}\rho\,\mathrm{d}X,
\end{multline}
where the function $f(z)$ is approximated by \eqref{eq:f(z)}. It is worth noting that \eqref{modsol2}, and its approximated form \eqref{modsol3}, are valid for $r>0$. At $r=0$, the exact solution \eqref{eq:modsol} is simplified to
\begin{equation}\label{modsolzero}
	k_w(x,0)=\int\limits_{X=x_0}^{x-\delta}\int\limits_{\rho=0}^{3D}\frac{\nu_t(X)}{2U_0\phi(X,x)}\,\exp\left\{-\,\frac{\rho^2}{4\phi(X,x)}\,-\psi(X,x)\right\}\bigg(\frac{\partial U(X,\rho)}{\partial \rho}\bigg)^2\,\rho\,\mathrm{d}\rho\,\mathrm{d}X.
\end{equation}

\bibliographystyle{jfm}
\bibliography{reference}

\end{document}